\documentclass[aps,prl,twocolumn,preprintnumbers,superscriptaddress,amsmath,longbibliography]{revtex4}

\usepackage{amssymb}
\usepackage{txfonts}
\usepackage{graphicx}
\usepackage{subfigure}
\usepackage{dcolumn}
\usepackage{bm}
\usepackage{color}
\usepackage{upgreek}

\newcommand{\ket}[1]{\left| #1 \right\rangle}

\begin{document}


\title{On-chip quantum interference between the origins of a multi-photon state}

\author{Lan-Tian Feng}
\affiliation
{CAS Key Laboratory of Quantum Information, University of Science and Technology of China, Hefei 230026, China.}
\affiliation{CAS Synergetic Innovation Center of Quantum Information $\&$ Quantum Physics, University of Science and Technology of China, Hefei 230026, China.}
\affiliation{Hefei National Laboratory, University of Science and Technology of China, Hefei 230088, China.}
\author{Ming Zhang}
\affiliation{State Key Laboratory for Modern Optical Instrumentation, Centre for Optical and Electromagnetic Research,
Zhejiang Provincial Key Laboratory for Sensing Technologies, Zhejiang University, Zijingang Campus, Hangzhou
310058, China.}
\author{Di Liu}
\author{Yu-Jie Cheng}
\author{Guo-Ping Guo}
\affiliation
{CAS Key Laboratory of Quantum Information, University of Science and Technology of China, Hefei 230026, China.}
\affiliation{CAS Synergetic Innovation Center of Quantum Information $\&$ Quantum Physics, University of Science and Technology of China, Hefei 230026, China.}
\affiliation{Hefei National Laboratory, University of Science and Technology of China, Hefei 230088, China.}
\author{Dao-Xin Dai}
\affiliation{State Key Laboratory for Modern Optical Instrumentation, Centre for Optical and Electromagnetic Research,
Zhejiang Provincial Key Laboratory for Sensing Technologies, Zhejiang University, Zijingang Campus, Hangzhou
310058, China.}
\author{Guang-Can Guo}
\affiliation
{CAS Key Laboratory of Quantum Information, University of Science and Technology of China, Hefei 230026, China.}
\affiliation{CAS Synergetic Innovation Center of Quantum Information $\&$ Quantum Physics, University of Science and Technology of China, Hefei 230026, China.}
\affiliation{Hefei National Laboratory, University of Science and Technology of China, Hefei 230088, China.}
\author{Mario Krenn\footnote[1]{mario.krenn@mpl.mpg.de}}
\affiliation{Max Planck Institute for the Science of Light (MPL), Erlangen, Germany.}
\affiliation{Department of Chemistry $\&$ Computer Science, University of Toronto, Toronto, Canada.}
\affiliation{Vector Institute for Artificial Intelligence, Toronto, Canada.}
\author{Xi-Feng Ren\footnote[2]{renxf@ustc.edu.cn}}
\affiliation
{CAS Key Laboratory of Quantum Information, University of Science and Technology of China, Hefei 230026, China.}
\affiliation{CAS Synergetic Innovation Center of Quantum Information $\&$ Quantum Physics, University of Science and Technology of China, Hefei 230026, China.}
\affiliation{Hefei National Laboratory, University of Science and Technology of China, Hefei 230088, China.}

\begin{abstract}
Path identiy induces a broad interest in recent years due to the foundation for numerous novel quantum information applications. Here, 
we experimentally demonstrate quantum coherent superposition of two different origins of a four-photon state, where multi-photon frustrated interference emerges from the quantum indistinguishability by path identity. The quantum state is created in four probabilistic photon-pair sources on one integrated silicon photonic chip, two combinations of which can create photon quadruplets. Coherent elimination and revival of distributed four-photons are fully controlled by tuning phases. 
The experiment gives rise to peculiar quantum interference of two possible ways to create photon quadruplets rather than interference of different intrinsic properties of photons. Besides many known potential applications, this new kind of multi-photon nonlinear interference enables the possibility for various fundamental studies such as nonlocality with multiple spatially separated locations.

\end{abstract}
\pacs{}
\maketitle

\section*{1. INTRODUCTION}
In 1994, Herzog, Rarity, Weinfurter and Zeilinger demonstrated a remarkable quantum interference effect \cite{herzog1994frustrated}. Like induced coherence without induced emission \cite{wang1991induced}, they overlapped the two paths of emerging photon pairs from two probabilistic photon-pair sources, in such a way that it cannot be distinguished -- not even in principle -- whether the pair has been created in the first or second source. As a result, the photons are in a coherent superposition of being created in either of the two sources. This generalization of superpositions of properties of photons to the mere origin of their creation has exciting consequences. They now tune a phase between the two crystals, and thereby manipulate the total amount of emerging photon pairs. 
This effect has been called frustrated quantum interference \cite{hochrainer2021quantum}, also known as the quantum version of one type of nonlinear interference \cite{chekhova2016nonlinear,ono2019observation,Luo2021}.


Frustrated pair-creation has been the enabling principle behind numerous emerging quantum technology. For example, it allows for a new form of infrared spectroscopy \cite{kalashnikov2016infrared} or terahertz quantum sensing  \cite{kutas2020terahertz} by exploiting visible light for which cheap and efficient photon detectors exist. So while this phenomena has been much investigated at the two-photon level, a multi-photon generalization has not even been proposed until 2017 \cite{krenn2017quantum, gu2019quantum}. There, taking inspiration of graph theory, it was discovered that the very particular alignment of paths in a four-crystal setup could lead to genuine four-photon interference that is not observable from two photons alone \cite{gu2019quantum}. The effect can be understood by a coherent superposition of two origins of a four-photon state that interfere.

Here we report the first experimental demonstration of this new multi-photon interference effect. In our four-photon experiment, the paths of the photons are overlapped in such a way that there are exactly two different possibilities of how each of the four detectors sees a single photon. Taking advantage of high quality integrated optics at a silicon chip, we align the photon paths in such a way that it cannot be distinguished whether the pairs have been created in the first or second possibility. In that way, the origin of the four photons is in a coherent superposition and we can observe constructive and destructive interference of four-photon systems that cannot be observed in the photon pairs themselves.

Our demonstration relies on a number of novelties compared to existing experiments. First, we extended the only other on-chip frustrated quantum interference experiment to multi-photon \cite{ono2019observation}, and the integrated pattern showed high scalabilty and would stimulate more new quantum optical experiments. Furthermore, we were able to demonstrate for the first time how two indistinguishable multipartite states beyond photon pairs can be generated experimentally using the concept of path identity \cite{hochrainer2021quantum}. This not only required the design of multiple sources of highly indistinguishable photon pairs, but also the precise mode matching between the different layers of photon quadruple generation.
Our work is the important part of path identity and lays the foundation for numerous novel applications such as quantum state generation, remote quantum metrology and observation of new nonlocal multiphoton interference effects \cite{hochrainer2021quantum}. Among others, the quantum interference effect we demonstrated here is the core experimental resource for a large class of new quantum state generation protocols that range from heralded entangled state generation to concepts for quantum simulation \cite{krenn2020conceptual,Ruiz2022}.



\section*{2. PHYSICAL INTERPRETATION}


Let's first look at the experiment by Herzog et al. \cite{herzog1994frustrated}, depicted in Fig. 1a. Each of the two nonlinear crystals can probabilistically create a photon pair. If all properties of the photon pair 
are identical, there is no way to distinguish from which crystal a pair was created. In that case, the two-photon state is

\begin{equation}
\begin{aligned}
\ket{\psi}=\frac{1}{\sqrt{2}}g\Big(\overbrace{\ket{a,b}}^\text{crystal II}+\overbrace{e^{i\varphi}\ket{a,b}}^\text{crystal I}\Big)=\frac{1}{\sqrt{2}}g\left(1+e^{i\varphi}\right)\ket{a,b},
\end{aligned}
\end{equation}
where $\ket{a,b}$ stands for a photon in path $a$ and one in path $b$, and $g$ is related with the pair creation probability. For clarity, we ignore the vacuum and higher-order terms.

\begin{figure}[t]
\centering
\includegraphics[width=6.0cm]{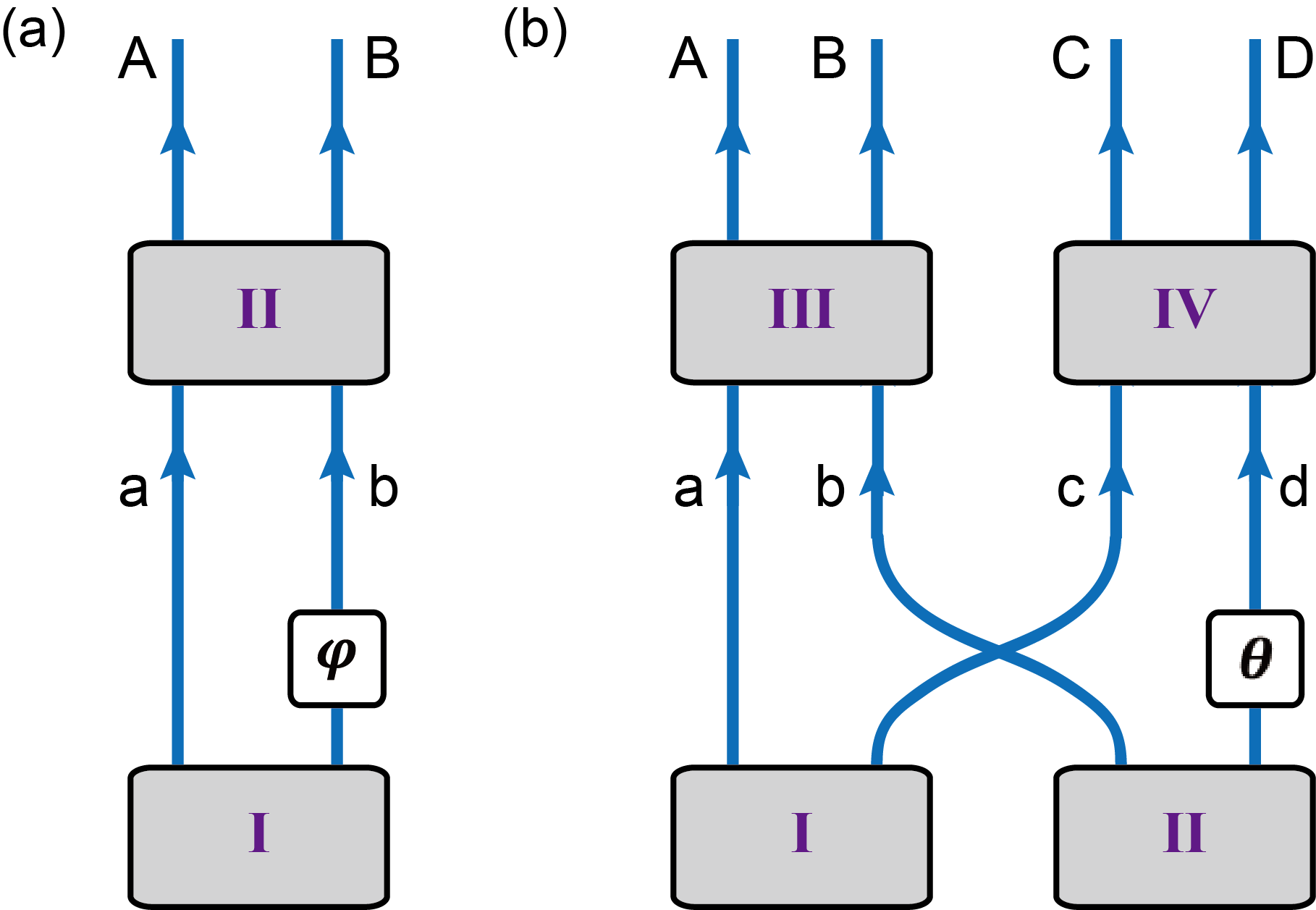}
 \caption {Quantum interference by indistinguishable origins. (a) Quantum interference between the origins of a two-photon state. 
(b) The multi-photon generalization. 
For the sake of simplification, the pump beams are not shown. In principle, the phases can also be added on the pump beams between crystals I, II and III, IV. 
}
 \label{TwoAndFourCrystals}
\end{figure}

The probability to detect a photon pair in $a$ and $b$ is $P_{a,b}\approx g^2 \left( 1 + \cos(\varphi)\right)$. With $\varphi=0$, constructive interference increases the number of photon pairs by a factor of 2 compared to the number of pair generated by a single crystal. 
With $\varphi=\pi$, the two possibilities for pair generation cancel each other. That means, even though the crystals would produce photons independently, 
no photon pair is created at all.

In our four-photon experiment, four photon-pair sources, I-IV, are used, as shown in Fig. 1b. Their paths are overlapped in such a way that detectors A-D only see photon quadruples if the source I \& II or III \& IV created a photon pair simultaneously. For small photon pair creation rate $g$, we can write the state as,

\begin{equation}
\begin{aligned}
|\psi\rangle&=\frac{1}{2}g\Big(\overbrace{|a,b\rangle}^\text{crystal III}+\overbrace{|c,d\rangle}^\text{crystal IV}+\overbrace{|a,c\rangle}^\text{crystal I}+\overbrace{e^{i\theta}|b,d\rangle}^\text{crystal II}\Big)\\
&+\frac{1}{4}g^2\Big(\underbrace{|a,b,c,d\rangle}_\text{crystal III\&IV}+\underbrace{e^{i\theta}|a,b,c,d\rangle}_\text{crystal I\&II}\Big) + \dots\\
&=\frac{1}{2}g\Big(|a,b\rangle+|c,d\rangle+|a,c\rangle+e^{i\theta}|b,d\rangle\Big)\\
&+\frac{1}{4}g^2\left(1+e^{i\theta}\right)|a,b,c,d\rangle+ \dots
\end{aligned}
\label{eq:generatorEq}
\end{equation}
where 
$\ket{a,b,c,d}$ stands for a photon quadruple in the detectors A-D. 
Again, by introducing a phase $\theta$ between the origins, we can observe constructive and destructive interference of the four-photon creation processes. For example, if the phase is $\theta=\pi$, no four-photon state can be observed in the detectors A-D, while each of the individual crystals continues to create photon pairs.

\section*{3. EXPERIMENTAL SETUP AND RESULTS}


\begin{figure*}[t]
\centering
\includegraphics[width=16.0cm]{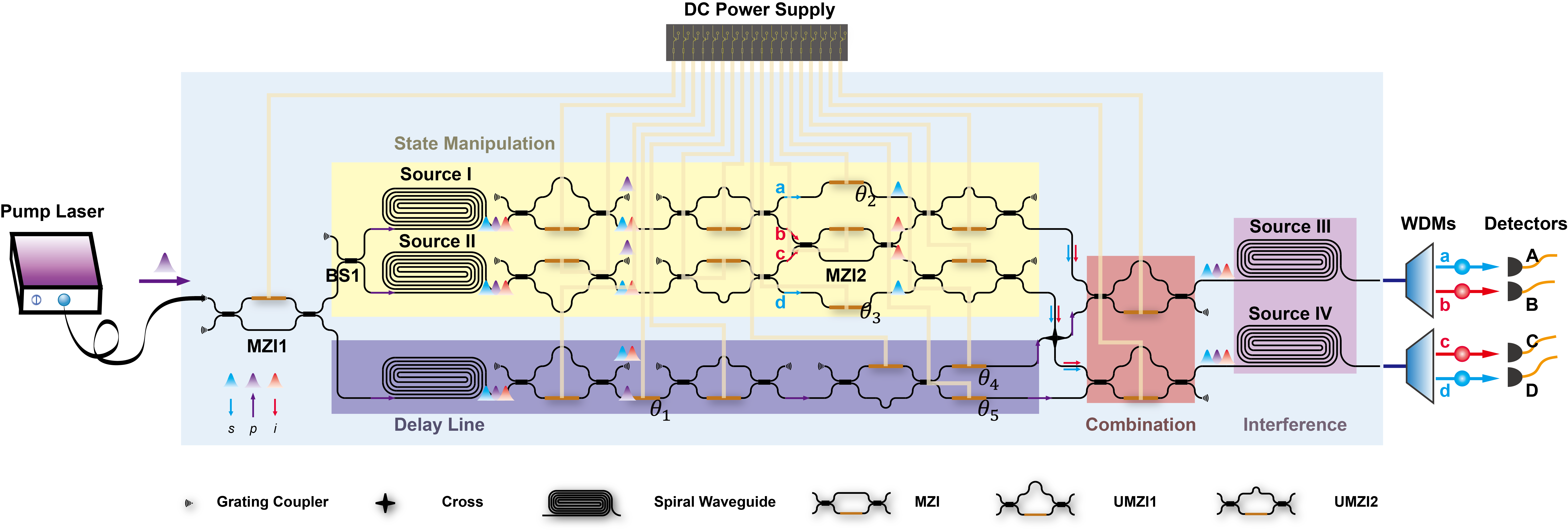}
 \caption {Experimental setup. 
Four spiral waveguide sources produce photon quadruples (Source I and Source II can produce a four-photon state as well as Source III and IV). 
A tunable MZI (MZI2) is used to change from two-photon interference to four-photon interference. 
MZI, Mach-Zehnder interferometer; UMZI, unbalanced Mach-Zehnder interferometer; WDMs: wavelength-division multiplexers.}
\label{fig:ChipLayout}
\end{figure*}

\begin{figure}[t]
\centering
\includegraphics[width=7cm]{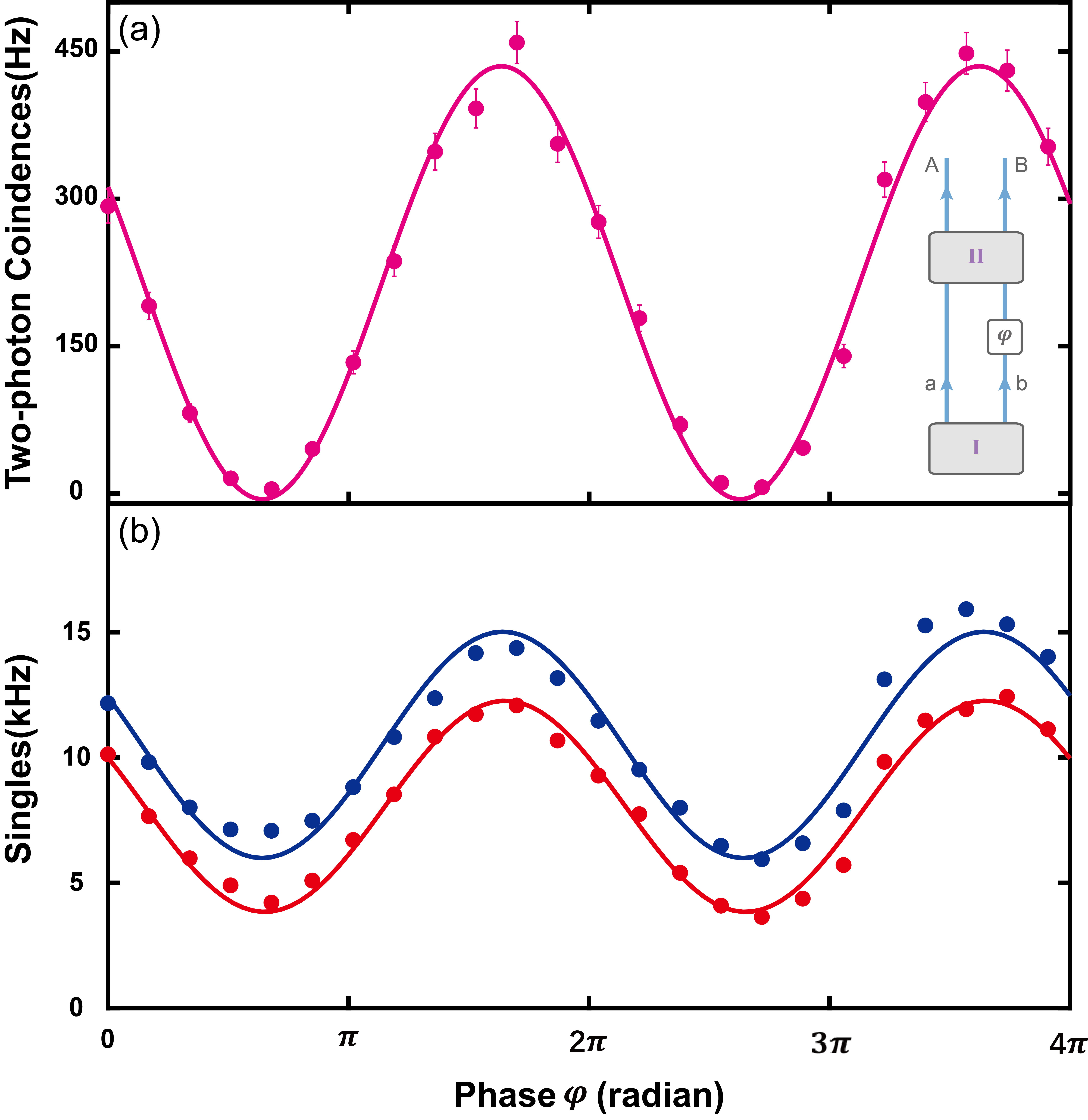}
\caption {Interference results with pump power 1.81 mW. Four-photon interference results (a) and two-photon fluctuation counts (b) when varying the phase $\theta_3$. 
Four-photon interference results (c) and two-photon fluctuation counts (d) when varying the phase $\theta_5$. Difference in two-photon counts is due to the imbalance in coupling and detection efficiency of different channels.}
\label{fig:my_plot3}
\end{figure}

\begin{figure}[t]
\centering
\includegraphics[width=7cm]{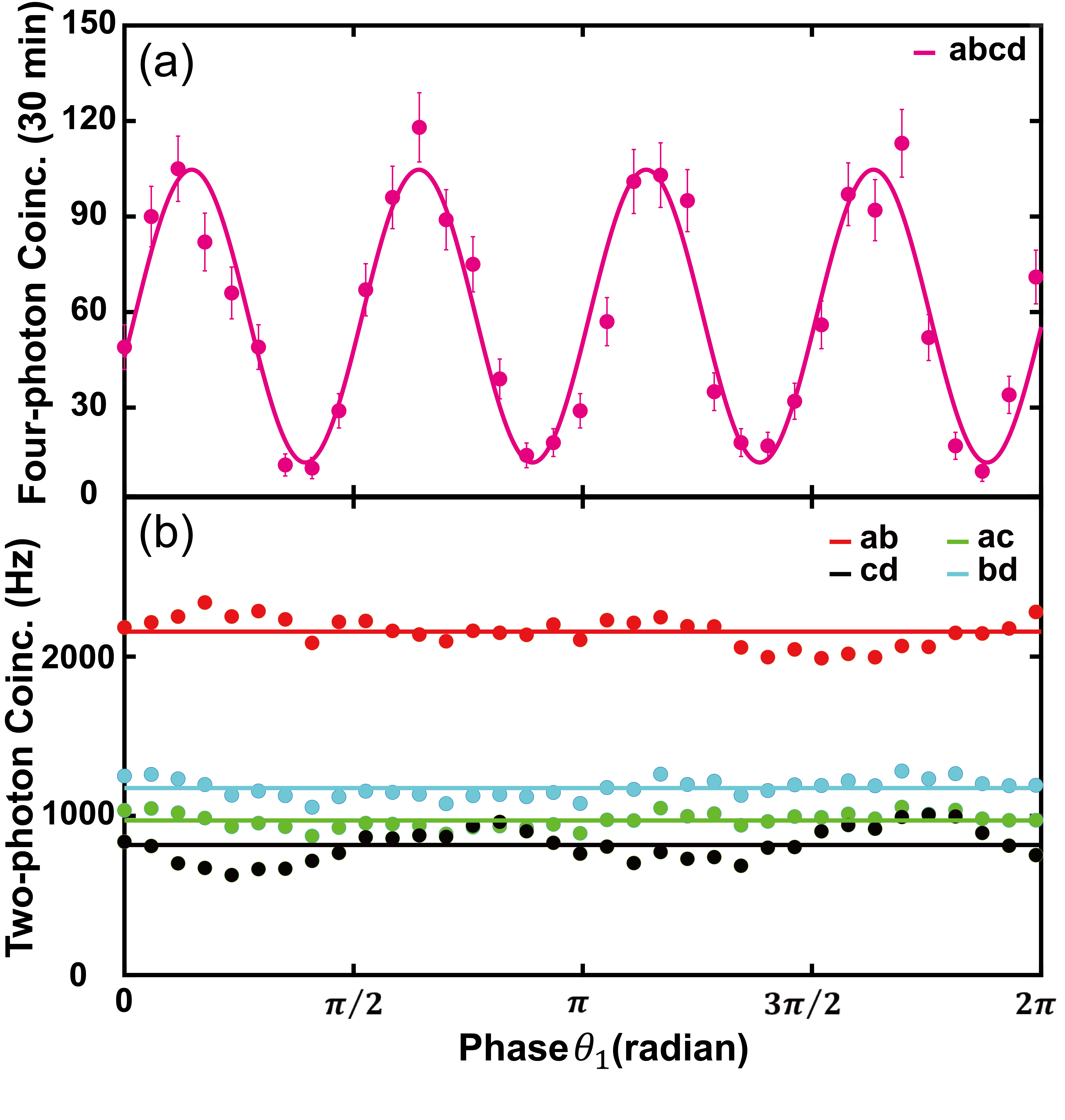}
\caption {Interference results with pump power 0.98 mW. Four-photon interference results (a) and two-photon fluctuation counts (b) when varying the phase $\theta_1$. 
}
\label{fig:fourPhotonResults}
\end{figure}

High phase-stability and mode matching among all the sources is quite important for our experiment, thus employing bulk optics or fiber optics will be challenging. The integrated photonic devices enable localization and manipulation of photons at micro/nano scale, thus greatly improve the stability and scalability \cite{wang2020integrated}. We explore this multi-photon interference effect using an integrated silicon photonic chip, and the conceptual layout is shown in Fig. \ref{fig:ChipLayout}. Integrated silicon photonics is becoming a scalable platform for quantum information processing in recent years 
\cite{silverstone2016silicon,paesani2019generation,zhang2019generation,elshaari2020hybrid,feng2020progress}. 
In the experiment, we pump four integrated photon-pair sources coherently and superpose different origins of the four-photon state to observe the multi-photon frustrated interference. All phases are modulated by on-chip heaters.


Specifically, a 200 GHz bandwidth pulsed laser centred at 1550.11 nm acts as the pump. It is coupled into the chip using grating couplers and split coherently into two paths using a Mach-Zehnder interferometer configured to act as a 50:50 beam splitter (MZI1). The upper beam is split again at a 50:50 beam splitter (BS1) and acts as the pump for sources I and II, while the lower beam is delayed and subsequently pumps sources III and IV. Each photon-pair source (I-IV) is a 5 mm long single-mode spiral silicon waveguide. In the waveguides, two pump photons are annihilated, and signal-idler photon pairs are generated with the spontaneously four-wave mixing process \cite{feng2019}. In the experiment, we select the signal and idler photons with respective central wavelength of 1545.32 nm and 1554.94 nm and 100 GHz bandwidth (see Supplement 1, Fig. S3). After sources I and II, an unbalanced Mach-Zehnder interferometer (UMZI1) is used to filter out the pump beam. Subsequently, the signal and idler photons in both paths are then separated using another unbalanced Mach-Zehnder interferometer (UMZI2). 
In this way, we can achieve four photons in four paths as $|a,b\rangle$ and $|c,d\rangle$.
In paths $b$ and $c$, we place an MZI to act as a tunable beam splitter (MZI2). 
After MZI2, another pair of UMZI2 are used to combine signal and idler photons into one path.
The lower pump (in the \textit{Delay Line} region) is combined with the photons created in sources I and II using two UMZI1s (in the \textit{Combination} region). The overlap is temporally controlled such that it enters into source III and IV at the same time as the photon pairs created in sources I and II. 
Extra phases $\theta_1$, $\theta_2$, $\theta_3$, $\theta_4$ and $\theta_5$ are introduced to observe frustrated quantum interference effect. 
After photon pairs are coupled out from the chip, they are delivered into four single-photon detectors for detection and further time correlation analysis. More details about chip characterization are given in Supplement 1. 

By setting MZI2 into the setting where the incoming photons remain in their paths, we demonstrate the two-photon frustrated interference process. The results are given in Supplement 1 and confirm the full controllability and high stability of the device. With the same filtering condition, we set MZI2 to swap the paths of the photons in $b$ and $c$, resembling the configuration in Fig. \ref{TwoAndFourCrystals}b. We get a two-photon state expressed as
\begin{equation}
\begin{aligned}
|\psi\rangle_2&=\frac{1}{2}\Big(\overbrace{e^{i\theta_2}|a,c\rangle}^\text{Source I}+\overbrace{e^{i\theta_3}|b,d\rangle}^\text{Source II}\\
&+\overbrace{e^{2i(\theta_1+\theta_4)}|a,b\rangle}^\text{Source III}+\overbrace{e^{2i(\theta_1+\theta_5)}|c,d\rangle}^\text{Source IV}\Big).
\end{aligned}
\end{equation}
Correspondingly, the four-photon state is expressed as
\begin{equation}
\begin{aligned}
|\psi\rangle_4&=\frac{1}{4}\Big(\overbrace{e^{i(4\theta_1+2\theta_4+2\theta_5)}|a,b,c,d\rangle}^\text{Source III\&IV}+\overbrace{e^{i(\theta_2+\theta_3)}|a,b,c,d\rangle}^\text{Source I\&II}\Big)\\
&=\frac{1}{4}(e^{i(\theta_2+\theta_3)}+e^{i(4\theta_1+2\theta_4+2\theta_5)})|a,b,c,d\rangle.
\end{aligned}
\end{equation}
From these equations, if we change phases ($\theta_1$, $\theta_2$, $\theta_3$, $\theta_4$ and $\theta_5$), four-photon coincidence counts will increase or decrease, while two-photon coincidence counts will keep unchanged. This is a multiphotonic generalization of two-photon frustrated quantum interference.


We first record four-photon coincidence counts when varying the phase $\theta_3$, and the results are given in Fig. 3a. This is exact Fig.1b scheme, that is, varying the phase on one photon. The pump power before the chip is set as 1.81 mW. 
Based on Eq. (4), the interference fringe can be fitted with $1+V \sin[\pi(\varphi-\varphi_c)/T]$, where $V$ is the fringe visibility, $\varphi_c$ is the initial phase, and $T$ is the oscillation period. The fringe visibility $V$ is defined as $V=(d_{\rm{max}}-d_{\rm{min}})/(d_{\rm{max}}+d_{\rm{min}})$, where $d_{\rm{max}}$ and $d_{\rm{min}}$ are the maximum and minimum of the fitted data, respectively. 
The raw visibility of the four-photon coincidence fringe was estimated as 66.7\%$\pm$3.1\%, and the error bar was obtained from 500 Monte Carlo simulations assuming a Poissonian distribution of the measured counts (the same below). 
Two-photon coincidence counts with linear fitting are given in Fig. 3b, and they show very small fluctuations. 

Besides phase $\theta_3$, varying other phases on the silicon chip will also observe similar frustrated interference effect. We record four-photon coincidence counts when varying the phase $\theta_5$ as an example, and the results are given in Fig. 3c. At this point, we actually vary the phase of the pump laser of one nonlinear crystal. The fitting curve shows two-fold oscillation frequency and the raw visibility of the four-photon coincidence fringe was estimated as 70.7\%$\pm$2.2\%. Corresponding two-photon coincidence counts are given in Fig. 3d, and they are basically the same as the results in Fig. 3b. 

To improve the visibility of four-photon interference results, we lower the pump power to 0.98 mW to reduce the multi-photon noise. In this regime, the photon-pair generation rate of each source is estimated as $p=0.01$, corresponding to $g=0.1$. We record four-photon coincidence counts with integral time of each point 30 min when varying the phase $\theta_1$, and the results are given in Fig. 4a. At this point, we actually vary the phase of the pump laser of two nonlinear crystals (one quadruplet, sources III and IV). The fitting curve shows four-fold oscillation frequency, and the raw visibility of the four-photon coincidence fringe was estimated as 78.3\%$\pm$2.6\%. In addition to the multi-photon noise, the spectral purity of single photons and imbalance of the interference terms will also reduce the four-photon interference visibility \cite{ono2019observation,adcock2019programmable,Faruque:18,Llewellyn2020,Meyer2017}. According to our experimental conditions, that is, pump laser with 200 GHz bandwidth, signal-idler photons with 100 GHz bandwidth and photon-pair generation rate 0.01, the maximum achievable visibility is estimated to be 0.84. Further visibility improvement can be achieved by using photon-number resolving detectors and increasing the spectral purity of on-chip photon-pair sources with novel integrated parametric source designs \cite{paesani2020near,liu2020high}. 

In any case, only curve cd shows a slightly higher fluctuation (around 16.4\% visibility with interference curve fitting), which is presumed to be caused by imperfect on-chip filtering which leads to frustrated two-photon quantum interference. 
However, the fluctuation of two-photon coincidence counts cannot explain the very large oscillation and pattern of four-photon interference. These results prove that we observe a totally new type of constructive and destructive multi-photon quantum interference, which cannot be understood by the behaviour of local properties such as individual photon-pair productions (see SI for details).  



\section*{4. DISCUSSION AND CONCLUSION}

In contrast to the two-photon case, the multi-partite generalization of frustrated quantum interference demonstrated here has non-trivial consequences. This time not only the origin of the four-photon state but also each origin consists of two locations that could be spatially separated. 
In this network, for example, source III could be in main chip, and source IV in another chip. They can be easily separated with dozens of kilometers with chip-to-chip setting linked with commercial fibers \cite{wei2020}. Because the pump laser and single photons all pass through the fiber, phases are still stabilized (see Supplement 1 for details). This separation will promote many exciting foundational experiments, for example, investigating more time delays of the interference effects \cite{Ma2016quantum}.

To illustrate the uniqueness of this nonlocal quantum interference effect, let’s imagine that the phase $\theta$ is close to crystal IV, and crystals III and IV are separated by a large distance. In that case, if we change the phase from 0 to $\pi$, and observe a photon pair in C and D, then we know that there cannot be any photon pairs emitted in crystal III. Thereby, we can make statements about spatially separated, and intrinsically random processes. This nonlocal interference resembles some properties of entanglement (where we can make statements about spatially separated, intrinsically random processes too). The crucial difference here is that we do not have superpositions of properties of photons, but between their origins. Observing this new type of nonlocal interference is a interesting goal for future quantum experiments \cite{hochrainer2021quantum}. Our experiment lays a stepping stone by demonstrating that this new quantum interference can indeed be observed in experiments.

Beyond the immediate physical interest, our work also suggests advances in the context of quantum technology applications. Based on our work, nonlinear interferometer with more photons can be constructed, and any application related to the two-photon version can be expanded to more photons. Besides, the significance of a multiphoton nonlinear interferometer goes far beyond just expanding the number of involved photons.
Our experiment demonstrates the coherently creating and overlapping highly-distinguishable photon pairs from different sources. It is the basis of several new proposals such as special-purpose quantum computing scheme \cite{gu2019quantum} and photonic quantum states generation \cite{krenn2017quantum,krenn2017entanglement,krenn2020conceptual}. 
By introducing the ability to  amplify multi-photon states in a distributed, remote way, it could advance quantum technologies for quantum networks \cite{Agarwal2014}. In fact, numerous proof-of-concepts applications have used the idea of path identity to enhance quantum imaging, microscopy, or spectroscopy \cite{hochrainer2021quantum}. More advantages might be discovered in similar applications by exploiting the multi-photon interference effect realized here. More importantly, our work increases the capacity of integrated photonic circuits to process quantum information. The advantage in scalability of integrated photonic technologies enables hundreds or thousands of components to be integrated on the chip at the same time, which will greatly promote the realization of new quantum optics experiments and quantum technologies.


To conclude, we have experimentally observed constructive and destructive interference between two possible ways to create photon quadruplets on the silicon chip. 
This fundamentally multiparticle nonlinear quantum interference effect shows an intricate interconnection between quantum coherence, non-locality and entanglement, fueled by the concept of indistinguishability. We believe that our results are likely to spearhead and motivate 
follow-up studies relate to the foundation of quantum mechanics and quantum technology applications.

We thank Alba Cervera-Lierta for useful comments on the manuscript. M.K. thanks Armin Hochrainer, Mayukh Lahiri, Manuel Erhard and Anton Zeilinger for valuable discussions on the concept of path identity.

\textbf{Funding.}  
This work was supported by the National Natural Science Foundation of China (NSFC) (Nos. 62061160487, 12004373), the Innovation Program for Quatum Science and Technology (2021ZD0303200), the Postdoctoral Science Foundation of China (No. 2020M671860) and the Fundamental Research Funds for the Central Universities. This work was partially carried out at the USTC Center for Micro and Nanoscale Research and Fabrication. M.K. acknowledges support from the Austrian Science Fund (FWF) through the Erwin Schr\"odinger fellowship No. J4309.

\textbf{Disclosures.} The authors declare no conflicts of interest.

\textbf{Data availability.} Data underlying the results presented in this paper are not
publicly available at this time but may be obtained from the authors upon reasonable request.

\textbf{Supplemental document.} See Supplement 1 for supporting content.



\end{document}